\documentclass[preprint,12pt, a4paper]{elsarticle}

\usepackage{amssymb}
\usepackage{hyperref}
\usepackage{color}
\usepackage{subcaption, graphicx}
\usepackage{subcaption}
\journal{SoftwareX}

\begin{document}
\renewcommand{\labelenumii}{\arabic{enumi}.\arabic{enumii}}

\begin{frontmatter}



\title{~\textsc{MPAT}: Modular Petri Net Assembly Toolkit}


\author[label1]{Stefano Chiaradonna}
\author[label1]{Petar Jevti\'{c}}
\author[label2]{Beckett Sterner}
\address[label1]{School of Mathematical and Statistical Sciences\\ Arizona State University\\ Tempe, AZ}
\address[label2]{School of Life Sciences\\
Arizona State University\\
Tempe, AZ, Email: beckett.sterner@asu.edu}

\begin{abstract}

We present a Python package called Modular Petri Net Assembly Toolkit (\textsc{MPAT}) that empowers users to easily create large-scale, modular Petri Nets for various spatial configurations, including extensive spatial grids or those derived from shape files, augmented with heterogeneous information layers. Petri Nets are powerful discrete event system modeling tools in computational biology and engineering. However, their utility for automated construction of large-scale spatial models has been limited by gaps in existing modeling software packages. \textsc{MPAT} addresses this gap by supporting the development of modular Petri Net models with flexible spatial geometries.

\end{abstract}


\end{frontmatter}



\begin{table}[h!]
\begin{tabular}{|l|p{6.5cm}|p{6.5cm}|}
\hline
\textbf{Nr.} & \textbf{Code metadata description} & \textbf{MPAT} \\
\hline
C1 & Current code version & 0.0.1\\
\hline
C2 & Permanent link to code/repository used for this code version & \url{https://github.com/schiarad2354/Petri_Net} \\
\hline
C3  & Permanent link to Reproducible Capsule &\url{https://github.com/schiarad2354/Petri_Net}\\
\hline
C4 & Legal Code License   & Non-commercial license: CC BY-NC 4.0 \\
\hline
C5 & Code versioning system used & git\\
\hline
C6 & Software code languages, tools, and services used & Python, Spike\\
\hline
C7 & Compilation requirements, operating environments \& dependencies & Python 3.8, GeoPandas, Pandas, NumPy\\
\hline
C8 & If available Link to developer documentation/manual & \url{https://github.com/schiarad2354/Petri_Net} \\
\hline
C9 & Support email for questions & pjevtic@asu.edu \\
\hline
\end{tabular}
\caption{Code metadata.}
\label{codeMetadata} 
\end{table}


\section{Motivation and significance}

\subsection{Introduction}
Many physical and biological systems can be understood as processes involving varying types of interactions among spatially distributed modular components. Mathematical formalisms for discrete event systems (DES) have been widely used to represent the dynamics of complex systems and resilience modeling [1]. Petri Net models are a prominent approach to DES simulation that is especially suitable for modeling and analyzing systems characterized by concurrent and distributed processes using directed graphs [2,3,4]. Several software packages currently support Petri Net simulation modeling, including Spike and GPenSim [5,6]. However, to our knowledge, existing software does not provide functionalities for the automated construction of large-scale, modular Petri Net systems with flexible spatial geometries. \smallskip

\subsection{Heterogeneous information layers}
Automated construction of large-scale Petri Net models is an important stepping stone for building complex systems compromised of multiple heterogeneous models over space and time dependent on various information layers. For instance, when it comes to information layers, modelers in ecology leverage granular spatio-temporal terrain data for modeling microclimates [7] or leverage heterogeneous habitat spatial data for modeling animal geographic patterns [8]. Similarly, for modeling natural hazards, meteorological data, for example, is crucial for rapid urban risk mapping from floods [9], earthquakes [10], tornados [11], and hurricanes [12]. In addition, insurers also utilize meteorological data for predicting hail claims [13], improving agriculture risk assessment [14], developing weather insurance for specific crops [15], etc. In other instances, modelers may leverage multiple, heterogeneous information layers simultaneously. For example, [16] leveraged historical wind speed, social vulnerability, and power flow for power loss mitigation. Similarly, [17] utilized humidity, precipitation, soil moisture, solar radiation, wind direction, temperature, etc. To integrate the wide range of information layers that modelers may utilize, we introduce the Modular Petri Net Assembly Toolkit (MPAT) to automate the construction of modular, spatial Petri Net models to be executed using the Spike software.


\subsection{Overview of Petri Nets}
Precisely, a Petri Net can be defined as a tuple $PN = (P, T, F, M_{0})$, where $P$ is a set of places, $T$ is a set of transitions, and  $P$ and $T$ are disjoint sets. The function $F: (T \times P) \cup (P \times T) \rightarrow N$ assigns a weight to each arc in the Petri Net. In particular, $M_{0}$ denotes the initial condition of the Petri Net as a function from the set of places to the positive integers. The arcs connect a place to a transition or a transition to a place. The places connected by an arc to a transition are called input places, and places with an arc coming from a transition are called output places. Arcs can be assigned weights with non-negative integer values. A transition becomes enabled when the number of tokens in the input place is at least the arc's weight. When the transition fires, it will consume a total number of tokens matching the arc weight from each input place and produce a total number of tokens matching the weight of each arc connecting to an output place. The distribution of tokens over the place represents a configuration or marking of the net. 

\indent~Petri Nets continue to be a popular modeling choice for a wide range of researchers in varying domains. As a testament to the widespread adoption of Petri Nets, a Google Scholar search using the keyword 'Petri Net' (accessed on July 1, 2024) returned approximately $224,000$ articles. In fields where spatial and temporal considerations are crucial, Petri Nets have proven particularly effective. For instance, they feature prominently in modeling wildfire propagation across heterogeneous spatial grids ($41,200$ articles), analyzing biochemical processes within heterogeneous cells for medical applications ($8,490$ articles), studying multi-scale environmental systems in ecology ($1,840$ articles), and modeling emergency resource allocation in urban settings ($1,260$ articles). Recent studies have also showcased their utility in epidemiology [18, 18, 20], resilience engineering [21], and computer engineering, particularly in spatial networking protocols for parallel and distributed systems [22, 23]. Despite this versatility of use of Petri Nets, to our knowledge, there is no assembly tool that can take different information layers relevant to an individual Petri Net and give their internal interconnectivity assessment of large-scale Petri Net. Typically, the only choice available to modelers is to use Colored Petri Nets.




\indent~Colored Petri Nets are popular for modeling spatial systems with a high level of symmetric structure because they simplify the model representation in terms of stacks of repeated components distinguished by different color labels [24, 25, 26, 27]. However, Colored Petri Nets also have limitations. They require considerable expertise in accurately modeling them, which can be a barrier for new users or those without a strong background in Petri Net methodology. Moreover, the benefits of using colored Petri Nets fall short when representing large models with heterogeneous structures and parameter settings [28], e.g. for systems with variable spatial grids and local processes. In particular, Colored Petri Nets typically model systems where components are repeated in a symmetric manner, often without spatial differentiation. For instance, in a spatial model where different grid cells have distinct environmental conditions, infrastructure levels, or economic activities, Colored Petri Nets struggle to differentiate these spatial entities effectively. This is because they tend to assume the same internal parametrized structure of the Petri Net model, i.e. the places and transitions for each grid cell. Colored Petri Nets require uniformity in the structure and behavior of their components, which means they cannot naturally encode large-scale spatial grids where each unit has unique characteristics or interactions. Automated model assembly is therefore important for scaling beyond small-scale grids, such as for a handful of places representing whole countries [18]. The challenges of Petri Net modeling are further underscored by the plethora of diverse software tools available for Petri Net modeling.



\subsection{Petri Net software tools}
\indent~There a number of published, open-source Petri Net simulation tools. Most rely on a graphical user interface (GUI) to construct and simulate Petri Net models, such as COSMOS [29], CPN Tools [30], PNetLab [31], ePNK [32], IOPT-Tools [33], WoPeD [34], ITS-Tools [35], and most recently Snoopy [36]. Other tools provide programming interfaces in languages such as MatLab [6,37] or Python [38]. However, many of these open-source tools are no longer under active development, or lack built-in features required to easily specify and modify large-scale systems. Additionally, none of the above have capabilities to run Petri Net models on a server, which can conserve user resources and accelerate the simulation process [5]. We therefore chose to focus on augmenting the Petri Net tool, Spike [5], which has been recognized for its efficiency and scalability [39, 40, 41, 42]. Our software package is also designed in a modular way to enable future modifications to be compatible with other toolkits.

\indent~\textit{Spike} is a command-line tool designed for continuous, stochastic, and hybrid simulation of Petri Nets [5]. It supports server-based operations, ensuring the reproducibility of models and configurations. Spike also supports Systems Biology Markup Language files [5], allowing for greater flexibility in modeling complex networks in a standard format for future integration and analysis. \textit{Systems Biology Markup Language} (SBML) is a file format designed for representing computational models in a declarative manner, allowing for seamless exchange between different software systems [43]. The exchange allows for wider utility in a variety of domains, such as modeling cellular systems [44], biochemical reactions [45], disease modeling [46], and Boolean networks with Petri Nets [47] (Another, related markup language, Petri Net Markup Language (PNML), does not appear to have been supported since 2015 [48, 49]). In addition, Spike has a command line feature that, unlike similar tools such as SNOOPY [36], is fully functional, making it suitable for automating multiple simulation experiments. \smallskip

\indent~However, to effectively utilize Spike, users must be proficient in manually modeling Petri Nets in either the SBML or Spike's internal format language, Abstract Net Description Language (ANDL). These formats are not commonly used or known outside of specialized scientific subfields and have their own intricacies and complexities. SBML moreover is based on XML, which is a widely used and mature standard but generally unfamiliar to students and researchers in the natural and social sciences. For a broader audience, then, there is clear value in facilitating the design of large-scale Petri Nets based on more commonly used file formats, such as shapefiles and CSV. This is particularly important for integration and interoperability of diverse datasets. To our knowledge, there is no assembly tool that can take heterogeneous information layers relevant to an individual Petri Net and give their internal interconnectivity assessment of large-scale Petri Net.

\subsection{Contributions}
\begin{figure}[t!]
\hspace*{-1.5cm}
    \centering
    \includegraphics[scale = 0.45]{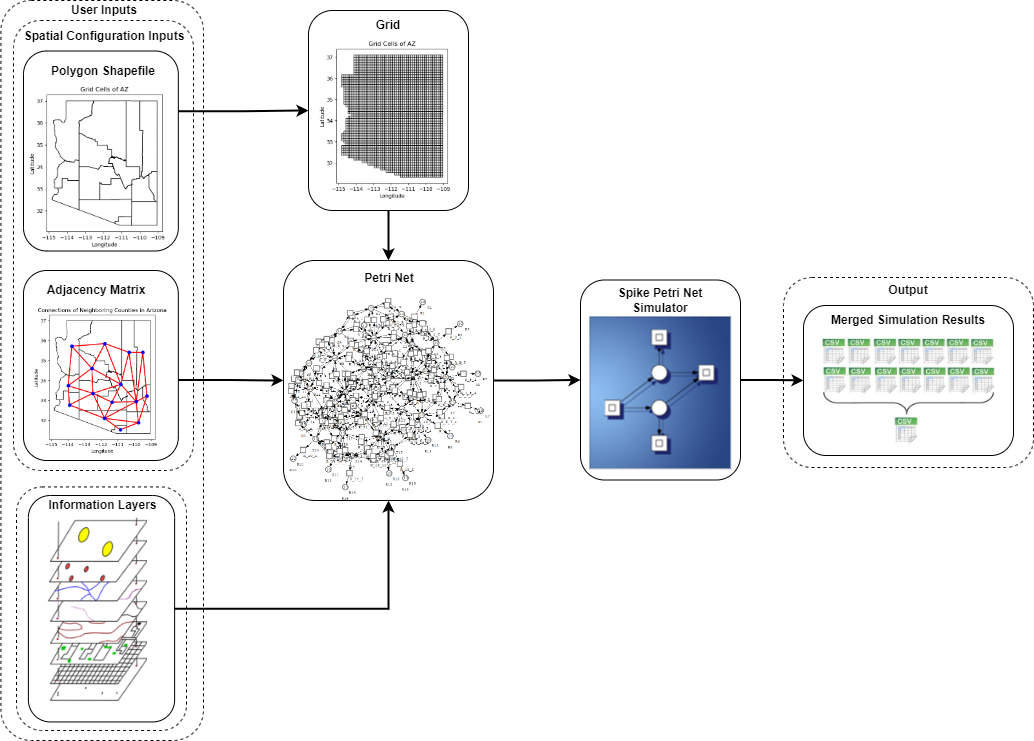}
    \caption{The proposed tool for modular assembly of large-scale Petri Nets.}
    \label{fig:proposed_tool}
\end{figure}

Specifically, we present a Python-based package, \textsc{MPAT}, that facilitates the modular assembly of large-scale Petri nets, empowering users to model extensive spatial grids with greater efficiency and ease. Rather than creating another Petri Net simulator from scratch, this package expands the utility of the existing tool Spike. The aim of~\textsc{MPAT} is to bridge the gap between the theoretical complexity of assembling Petri Net models and their practical application, making them more accessible and usable for a broader range of spatial modeling tasks. To our knowledge, no other software package provides these capabilities for the assembly of Petri Nets at this scale. The proposed toolkit for modular assembly of large-scale Petri Nets is shown in Figure~\ref{fig:proposed_tool}. \smallskip

\indent~In particular, the contributions of this research are fourfold:
\begin{itemize}
    \item[1.]~The proposed software toolkit takes a shapefile and generates CSV file describing a grid of the desired size of the spatial Petri Net model with adjacent patches (or grid cells) connected. This streamlines the process of creating and managing complex models without needing to manually configure each connection.
    \item[2.]~As another option, the software takes input from a CSV file of the adjacency matrix of the patches (or grid cells) and outputs the spatial structure of the Petri Net model, complete with interconnected places and transitions.
    \item[3.]~The proposed software can take in heterogeneous, individual information layers, such as vegetation, wind speed, humidity, etc. in a vector format. By integrating these diverse information layers, the software enables users to fine-tune Petri Net parameters at more granular spatio-temporal levels not currently available in existing Petri Net software tools.
    \item[4.]~The software streamlines efficient parameter search across heterogeneous parameter spaces by consolidating Petri Net results into a unified CSV file. This supports comprehensive analysis and comparison of various model configurations.
    \end{itemize}

\indent~The remainder of this paper is structured as follows: Section~\ref{description} provides an introduction to the structure and features of~\textsc{MPAT}. In Section~\ref{examples}, we present two use cases that demonstrate the validity and the utility of~\textsc{MPAT}, while Section~\ref{impact} describes the software's impact and broader implications. Finally, in Section~\ref{conclusion}, we share our insights into the new software toolkit and future directions in the concluding section.

\section{Software description}\label{description}
In this section, we examine the proposed tool, \textsc{MPAT}. First, we outline the software architecture, followed by a detailed exploration of its functionality and the formats of its input and output.


\subsection{Software architecture}
\begin{figure}[t!]
\hspace*{-3cm}
    \centering
    \includegraphics[scale = 0.33]{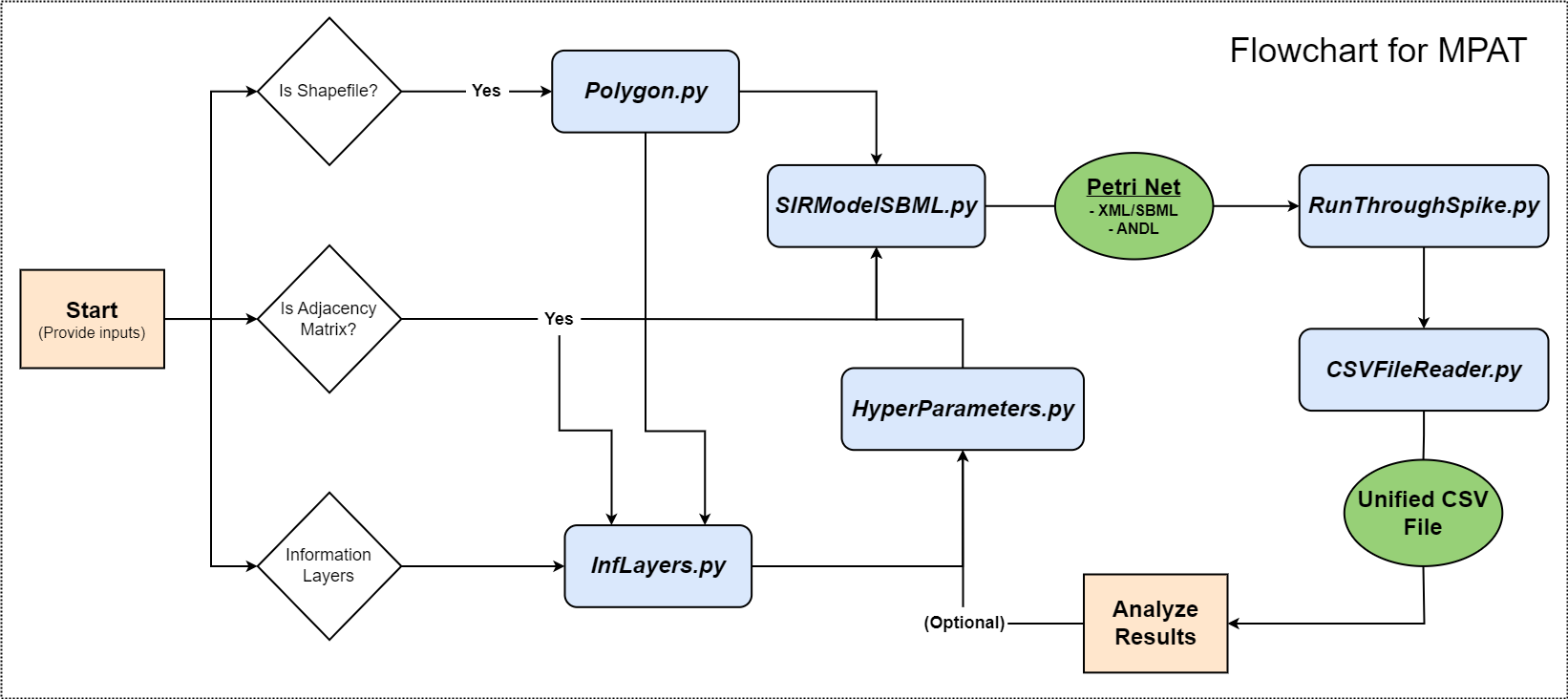}
    \caption{The flowchart of MPAT functions. The triangles indicate the initial input arguments. The blue rectangles indicate the Python functions and the green circles indicate the output files.}
    \label{flowchart}
\end{figure}

\textsc{MPAT} has five primary Python source codes, specifically~\emph{Polygon.py}, \emph{InfLayers.py}, \emph{SIRModelSBML.py}, \emph{HyperParameters.py}, \emph{RunThroughSpike.py}, and \emph{CSVFileReader.py}. Figure~\ref{flowchart} provides a flowchart of the high-level overview of MPAT. Below, we describe each function in Figure~\ref{flowchart}.\smallskip

\indent~To start, \emph{Polygon.py} manages the conversion of a shapefile into a GeoPandas dataframe, organizing grid cells into patches and linking them to their neighboring patches. Additionally, \emph{InfLayers.py} handles the information layers in vector format. \emph{SIRModelSBML.py} generates the corresponding Petri Net configuration model while \emph{HyperParameters.py} handles the parameter space for the model. Next, \emph{RunThroughSpike.py} handles the Petri Net simulations by running each configuration model through the Petri Net simulator, Spike [50]. To avoid any compatibility difficulties, the user sets the path to Spike. Finally, the output of the simulation results is handled by \emph{CSVFileReader.py} to facilitate the accessibility of the results and an efficient way for analysis.

 \subsection{Software functionalities}
In this section, we describe the major functionalities of \textsc{MPAT} (see Figure~\ref{flowchart}). To begin, the user inputs the choice of a shape file for a geographic area or the adjacency file as a CSV file of predefined patches. If it is a shapefile, \emph{Polygon.py} generates a grid of patches of the user's desired size and outputs the GeoPandas data frame of the grid and adjacency matrix. On the other hand, if the adjacency matrix file is already defined by the user, then we proceed directly to the Petri Net modeling component of the toolkit. Additionally, the user input of information layers flows into \emph{InfLayers.py} for linking with the grid of patches from \emph{Polygon.py} or adjacency matrix. Given the adjacency matrix either from \emph{Polygon.py} or the user, \emph{SIRModelSBML.py} generates the Petri Net model with desired input parameters, such as the initial number of tokens/markings for each patch and the arc weights between the patches. For the initialization of one instance of the parameters, such as tokens/markings, the user can import a CSV file with the corresponding name of the place (transition) and the number of tokens/markings (arc weight). For places that are not specified, the default token/marking is zero, and the default arc weight is one for transitions. If no CSV is imported, then the default settings of 100 tokens/markings in the first place of the order of the Petri Net are assigned, and the default arc weight of the transitions is one. The output is the corresponding Petri Net model in XML/SBML and ANDL file formats. Generating the SBML and ANDL files uses Python's built-in xml.etree submodule. These files are the reference configuration files for parallelizable or multi-scale modeling for user-defined hyperspace of parameter values in \emph{HyperParameters.py}. Given the Petri Net model files, \emph{RunThroughSpike.py} runs each file through the Petri Net simulator, Spike. The output of \emph{RunThroughSpike.py} is a collection of CSV files of the resulting simulation results for each simulation run. Because of the unwieldy number of CSV files produced by executing multiple runs, \emph{CSVFileReader.py} compiles all of the CSV files into a single, creating an automatic directory and providing a basic analysis of the results. Since the files are in CSV format, they can be exported to a variety of software applications by the user.


\subsection{Input and output formats}
\textsc{MPAT} has three primary input options. The first option is a shapefile of the geographic region. A shapefile, often denoted by the common extension .shp, serves as the standard format for representing spatial data or features, encompassing polygons. In the shapefile, the polygon is a fundamental unit of spatial representation. The second input option is a CSV file of the adjacency matrix of patches (or grid cells). The third is an optional CSV file of initialized parameters, such as tokens/markings and arc weights, for one instance of the Petri Net model. \smallskip

\indent~An intermediate output is the corresponding Petri Net model in ANDL (.andl) and SBML (.xml) files. After executing the model runs in the Petri Net simulator, Spike, \textsc{MPAT} compiles the collection of CSV files into a merged CSV output describing the Petri Net simulation results for analysis and visualization. We now proceed to illustrate the functionality and utility of~\textsc{MPAT}.

\section{Illustrative examples}\label{examples}

In this section, we provide two examples to demonstrate the utility of the proposed tool. The first is validation with a percolation model describing the spreading of forest fires. The second demonstrates the assembly capabilities of the tool in the context of epidemiological models for United States counties. The examples are found under the examples folder in the~\textsc{MPAT} GitHub repository.

\subsection{Validation with a percolation model of forest fire spreading}
\begin{figure}[t!]
\hspace*{-2cm}
    \centering
    \includegraphics[scale = 0.45]{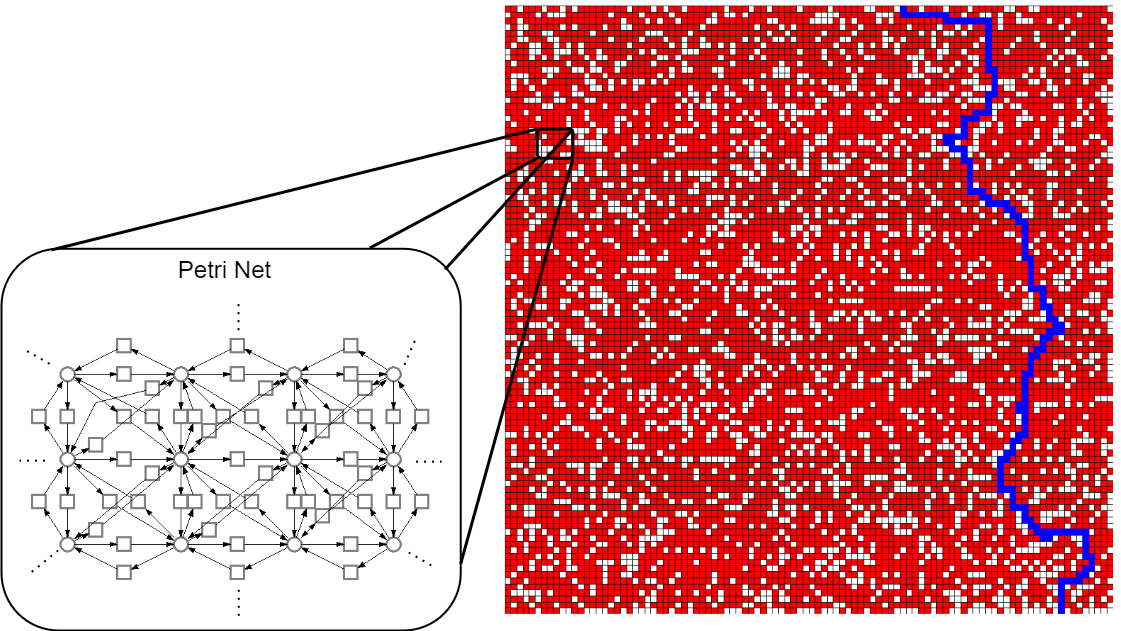}
    \caption{Visualization of one instance of the fire spread model with Petri Net representation on $100 \times 100$ lattice with probability $p = 0.41$ with Moore neighborhood. The white patches have the tree that is alive and the red patches have the tree on fire. The blue path is the percolation cluster.}
    \label{fig:percolationpath}
\end{figure}

In this example, we validate the results of the \textsc{MPAT} tool with those of a percolation model on a two-dimensional square lattice using the Moore neighborhood (i.e., chess queen adjacency between different patches). For motivation, consider the dynamics of a forest fire spreading through a landscape. Here, trees are represented by patches on the lattice, and the fire spreads from one tree to its neighboring trees according to the rules of percolation theory.\footnote{For a comprehensive introduction to percolation theory, we refer the reader to [51].} Specifically, each patch has a tree that is either alive (white patch) or on fire (red patch) (see Figure~\ref{fig:percolationpath}).\footnote{For simplicity, we assume that each patch has only one tree.} Furthermore, the critical density (denoted as $p_{c}$) is the probability at which there's a continuous path of patches with trees on fire from one side of the lattice to the opposite side. Below this critical density, the fire has a low probability of dying out because there isn't a sufficient connected path of patches with trees on fire. At or above this density, the fire can percolate through the entire lattice. In particular, at a critical density of \( p_{c} \approx 0.41 \), the fire spreading behavior shifts from an extinction regime, where the fire eventually dies out, to an uncontrollable regime, where the fire continues to spread, forming large clusters of burning trees [52,53, 54] (see the blue path in Figure~\ref{fig:percolationpath}). This critical point marks the transition between different behaviors of the system, demonstrating the importance of understanding percolation thresholds in predicting and managing the spread of forest fires. By comparing the \textsc{MPAT} tool's results with those of the established percolation model, we can validate the tool's accuracy in simulating such large-scale, spatial phenomena.

\begin{figure}[t!]
\begin{subfigure}{.475\linewidth} \hspace*{-3cm}
  \includegraphics[scale = 0.4]{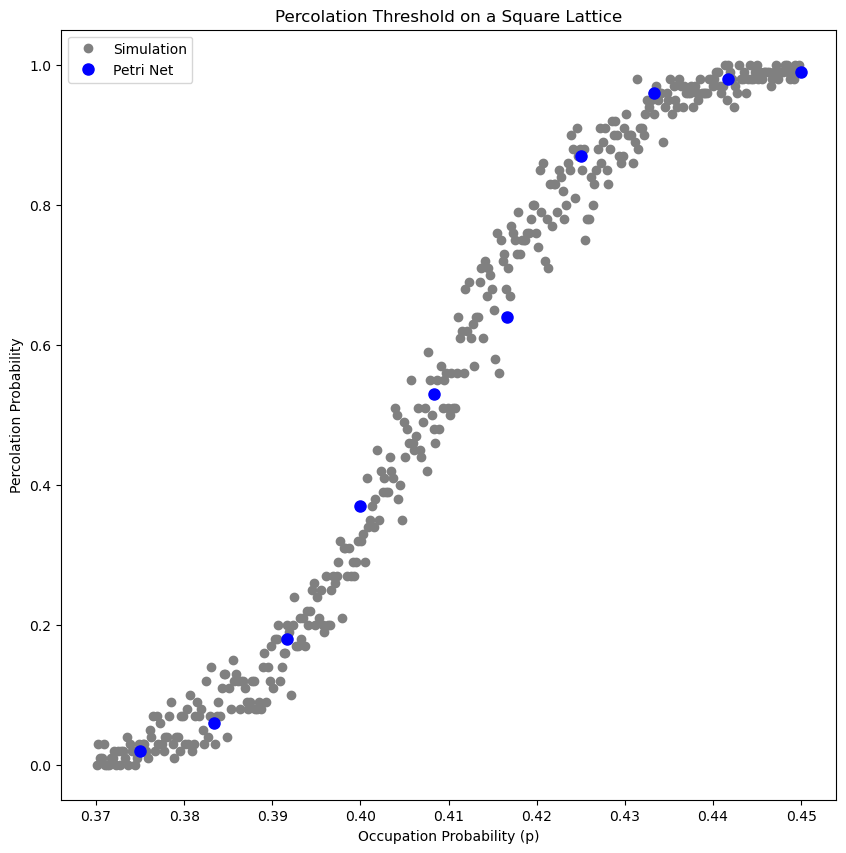}
  \caption{\hspace*{3cm}}
  \label{fig:pecthresh}
\end{subfigure}\hfill 
\begin{subfigure}{.475\linewidth}
  \includegraphics[scale = 0.4]{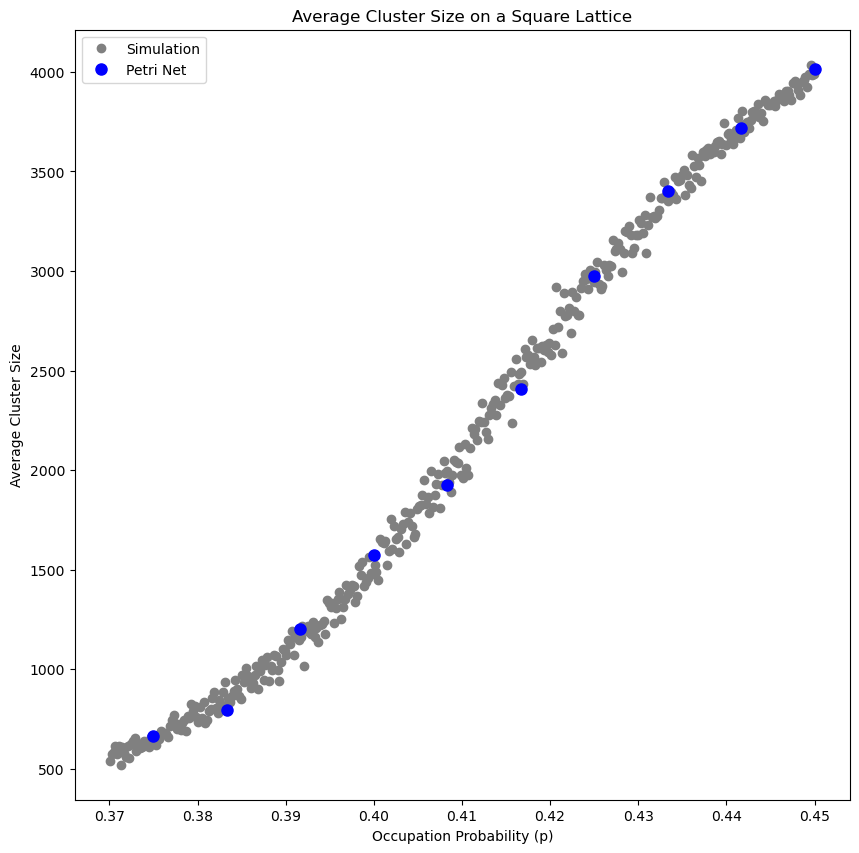}
  \caption{\hspace*{-3.5cm}}
  \label{fig:cluster}
\end{subfigure}
\caption{Visualizations of simulations (grey) and Petri Net results (blue) for the (a) percolation threshold and (b) average cluster size on a $100 \times 100$ lattice with a Moore neighborhood. The percolation threshold, at which a phase transition is present, occurs at $p_{c} \approx 0.41.$}
\label{fig:val}
\end{figure}

\indent~In particular, Figure~\ref{fig:val} compares the Petri Net model from \textsc{MPAT} with percolation model simulations, demonstrating the percolation threshold at \( p_{c} \approx 0.41 \). This result aligns with established findings in the mathematical literature [52, 53, 54]. Additionally, Figure~\ref{fig:cluster} illustrates the mean cluster size for different occupation probabilities, further validating~\textsc{MPAT}. 

\newpage \clearpage
\subsection{Large-scale, multi-patch epidemiological models}
\begin{figure}[t!]
    \centering
    \includegraphics[scale = 0.5]{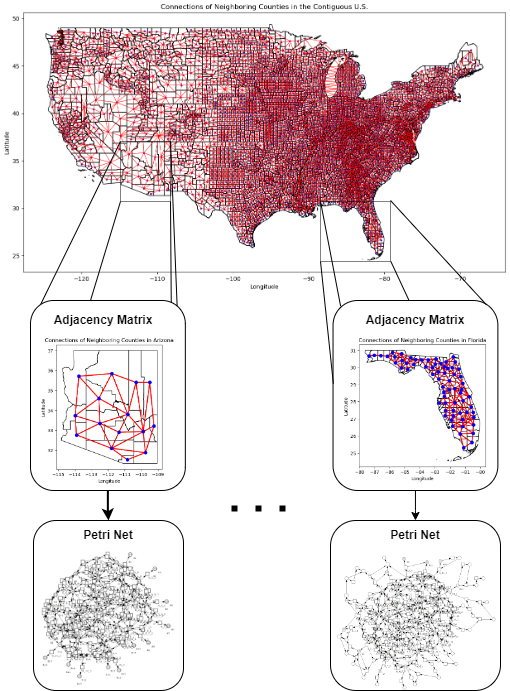}
    \caption{County level adjacency of the contiguous United States.\protect\footnotemark}
    \label{fig:counties}
\end{figure}

To illustrate the capabilities of \textsc{MPAT} for building large-scale spatial Petri Net models, we construct epidemiological models spanning all 3,066 counties in the contiguous United States. These models account for the interconnections with neighboring counties, utilizing the adjacency file from the U.S. Census Bureau [55] (see Figure~\ref{fig:counties}). For simplicity, we consider a standard Susceptible (S), Infected (I), and Recovered (R) model as a Petri Net with transitions between neighboring S and I counties.\footnote{For consistency, we construct the epidemiological models as Petri Nets following [18, 19] for each county.} Next, \textsc{MPAT} generates the corresponding SBML and ANDL files, which are then implemented in Spike. The output is a unified CSV file of simulations with different stylized parameters. The user can alter the initial parameters as desired for multiple scenarios. In terms of computing resources, the simulations were conducted on an Intel Xeon Gold 6226R CPU @ 2.90Ghz with 64 threads and 512 GB of memory Linux server. The resulting runtime was approximately 2 hours and 7 minutes.

\section{Impact}\label{impact}
The existing Petri Net simulation tools often rely on a GUI, making it cumbersome for large-scale assembly. For the tools that use a command-line interface, there is a critical lack of modular assembly of Petri Net models both at scale and in overcoming the specific language of the tool and Petri Net modeling structure. \smallskip

\indent~In addressing these challenges, \textsc{MPAT} is a Python-based modular assembly toolkit for Petri Net modeling. This development not only makes Petri Net modeling more accessible but also unlocks additional potential for Petri Net simulators, such as Spike. The \textsc{MPAT} toolkit allows for seamless integration of various data sources, information layers, and modeling requirements. By automatically generating a grid from shapefiles and establishing connections between adjacent patches, it simplifies the creation and management of complex spatial Petri Net models. The ability to input an adjacency matrix from a CSV file further enhances the flexibility and ease of use, enabling users to quickly define and visualize the spatial structure of their models. Additionally, \textsc{MPAT} streamlines the parameter search process across heterogeneous parameter spaces by consolidating results into a unified CSV file. This facilitates comprehensive analysis and comparison of different model configurations, making \textsc{MPAT} an invaluable tool for researchers and practitioners in the field of Petri Net modeling. \textsc{MPAT} enables users to process spatial data from various geographic regions more efficiently than ever before, paving the way for more sophisticated and granular analysis. \smallskip  \footnotetext{According to the source file for county adjacency [55], "in some instances, the boundary between two counties may fall within a body of water, so it seems as if the two counties do not physically touch.  These counties are included on the list as neighbors". This explains instances where counties are classified as neighbors near bodies of water. }

\indent~By leveraging the wide range utility of Python, \textsc{MPAT} makes complex and large-scale modeling of Petri Nets at varying spatial scales more manageable and user-friendly. The capabilities of \textsc{MPAT} using the SBML and ANDL formats allows for further extensions of this tool into the collection of Petri Net tools in the PetriNuts family of tools, such as Charlie [56], Patty [57], and Snoopy [36], as well as future integrations of machine learning algorithms for analysis (see e.g. [58,59]). \smallskip

\indent~In summary, \textsc{MPAT} is a powerful tool for the construction of large-scale, multi-patch Petri Nets for modular assembly. Rather than creating another software for Petri Net simulations, this tool expands the user-ability and modeling capabilities of an existing and commonly used Petri Net simulator, Spike. In doing so, \textsc{MPAT} helps reduce the learning curve of Spike but also broadens the use-cases of the modular assembly of Petri Nets into other areas, such as hydrology, management, and environmental science, where users may be unfamiliar with Petri Net modeling.


\section{Conclusions}\label{conclusion}
\textsc{MPAT} offers a user-friendly and flexible framework for the modular assembly of Petri Nets, paving the way for new research avenues of large-scale Petri Net modeling. The integration of heterogeneous vector data layers, such as population, vegetation, wind speed, and humidity, enhances modelers' abilities to parameterize Petri Net models at more granular spatio-temporal levels. By simplifying the user experience and boosting adaptability, the proposed software equips seasoned researchers and newcomers to Petri Nets with a new tool for large-scale spatial modeling and analysis. \smallskip

\indent~Future work may encompass several extensions of the software. The first is to expand it to the community of hybrid Petri Net models, which incorporate combinations of continuous and stochastic places and transitions. The second is to expand the pipeline for analysis to allow for a wider range of machine-learning resources, such as \textit{Scikit-learn} and \textit{SciPy}.

\section*{Acknowledgements}
\label{}
\emph{This work was supported under the National Institutes of Health grant DMS-1615879. The authors declare no conflicts of interest.}

\section*{References}
[1] S. Oliveira, A. B. Leal, M. Teixeira, Y. K. Lopes, A classification of
cybersecurity strategies in the context of discrete event systems, Annual
Reviews in Control 56 (2023) 100907. \\

[2] J. L. Peterson, Petri nets, ACM Computing Surveys (CSUR) 9 (3)
(1977) 223–252.\\

[3] W. Reisig, Elements of distributed algorithms: modeling and analysis
with Petri nets, Springer Science \& Business Media, 1998.\\

[4] W. Reisig, Petri nets: an introduction, Vol. 4, Springer Science \& Busi-
ness Media, 2012.\\

[5] J. Chodak, M. Heiner, Spike–reproducible simulation experiments with
configuration file branching, in: Computational Methods in Systems
Biology: 17th International Conference, CMSB 2019, Trieste, Italy,
September 18–20, 2019, Proceedings 17, Springer, 2019, pp. 315–321.\\

[6] R. Davidrajuh, Gpensim: a new Petri net simulator, InTech, 2010. \\

[7] M. R. Kearney, P. K. Gillingham, I. Bramer, J. P. Duffy, I. M. Maclean,
A method for computing hourly, historical, terrain-corrected microclimate anywhere on earth, Methods in Ecology and Evolution 11 (1)
(2020) 38–43. \\

[8] B. Jim enez-Alfaro, M. Chytry, L. Mucina, J. B. Grace, M. Rejm anek,
Disentangling vegetation diversity from climate–energy and habitat het erogeneity for explaining animal geographic patterns, Ecology and evo-
lution 6 (5) (2016) 1515–1526. \\

[9] L. Lin, C. Tang, Q. Liang, Z. Wu, X. Wang, S. Zhao, Rapid urban
flood risk mapping for data-scarce environments using social sensing
and region-stable deep neural network, Journal of Hydrology 617 (2023)
128758. \\

[10] Z. Jin, Y. Fialko, Finite slip models of the 2019 ridgecrest earthquake
sequence constrained by space geodetic data and aftershock locations,
Bulletin of the Seismological Society of America 110 (4) (2020) 1660–
1679. \\

[11] F. Valente, M. Laurini, Tornado occurrences in the united states: a
spatio-temporal point process approach, Econometrics 8 (2) (2020) 25. \\

[12] M. Sajjad, N. Lin, J. C. Chan, Spatial heterogeneities of current and
future hurricane flood risk along the us atlantic and gulf coasts, Science
of the total environment 713 (2020) 136704. \\

[13] L. Gao, P. Shi, Leveraging high-resolution weather information to pre-
dict hail damage claims: A spatial point process for replicated point
patterns, Insurance: Mathematics and Economics 107 (2022) 161–179. \\

[14] M. Soliman, N. K. Newlands, V. Lyubchich, Y. R. Gel, Multivariate cop-
ula modeling for improving agricultural risk assessment under climate
variability, Variance 16 (1) (2023). \\

[15] J. Bucheli, N. Conrad, S. Wimmer, T. Dalhaus, R. Finger, Weather
insurance in european crop and horticulture production, Climate Risk
Management 41 (2023) 100525. \\

[16] E. Boyle, A. Inanlouganji, T. Carvalhaes, P. Jevti´c, G. Pedrielli, T. A. Reddy, Social vulnerability and power loss mitigation: A case study of puerto rico, International Journal of Disaster Risk Reduction 82 (2022)
103357.\\

[17] V. Y. Chandrappa, B. Ray, N. Ashwatha, P. Shrestha, Spatiotemporal modeling to predict soil moisture for sustainable smart irrigation, Internet of Things 21 (2023) 100671. \\

[18] S. Connolly, D. Gilbert, M. Heiner, From epidemic to pandemic modelling, Frontiers in Systems Biology 2 (2022) 861562. \\

[19] C. Segovia, Petri nets in epidemiology, arXiv preprint arXiv:2206.03269
(2022). \\

[20] D. A. Zaitsev, T. R. Shmeleva, S. Gizurarson, Reenterable colored petri
net model of ebola virus dynamics, in: 2023 IEEE International Conference on Machine Learning and Applied Network Technologies (ICMLANT), IEEE, 2023, pp. 1–6. \\

[21] R. Yan, S. Dunnett, J. Andrews, A petri net model-based resilience
analysis of nuclear power plants under the threat of natural hazards,
Reliability Engineering \& System Safety 230 (2023) 108979. \\

[22] V. Kulagin, N. Muraviev, Software for modeling distributed systems
using the petri net apparatus, in: 2024 Wave Electronics and its Application in Information and Telecommunication Systems (WECONF),
IEEE, 2024, pp. 1–5. \\

[23] L. Capra, M. K¨ohler-Bußmeier, Modular rewritable petri nets: An efficient model for dynamic distributed systems, Theoretical Computer
Science 990 (2024) 114397. \\

[24] D. Gilbert, M. Heiner, F. Liu, N. Saunders, Colouring space-a coloured
framework for spatial modelling in systems biology, in: Application and
Theory of Petri Nets and Concurrency: 34th International Conference,
PETRI NETS 2013, Milan, Italy, June 24-28, 2013. Proceedings 34,
Springer, 2013, pp. 230–249. \\

[25] Q. Gao, F. Liu, D. Tree, D. Gilbert, Multi-cell modelling using coloured
petri nets applied to planar cell polarity, in: Proceedings of the 2nd
International Workshop on Biological Processes \& Petri Nets, satellite
event of PETRI NETS, Vol. 724, Citeseer, 2011, pp. 135–50. \\

[26] D. A. Zaitsev, T. R. Shmeleva, W. Retschitzegger, Spatial specification
of grid structures by petri nets, in: Micro-Electronics and Telecommunication Engineering: Proceedings of 4th ICMETE 2020, Springer, 2021,
pp. 253–263. \\

[27] D. A. Zaitsev, T. R. Shmeleva, B. Pr¨oll, Spatial specification of hypertorus interconnect by infinite and reenterable coloured petri nets, International Journal of Parallel, Emergent and Distributed Systems 37 (1)
(2022) 1–21. \\

[28] J. C. Carrasquel, I. A. Lomazova, A. Rivkin, Modeling trading systems
using petri net extensions., in: PNSE@ Petri Nets, 2020, pp. 118–137. \\

[29] P. Ballarini, B. Barbot, M. Duflot, S. Haddad, N. Pekergin, Hasl: A new
approach for performance evaluation and model checking from concepts
to experimentation, Performance Evaluation 90 (2015) 53–77.\\

[30] M. Westergaard, L. M. Kristensen, The access/cpn framework: A tool
for interacting with the cpn tools simulator, in: International Conference
on Applications and Theory of Petri Nets, Springer, 2009, pp. 313–322. \\

[31] F. Basile, C. Carbone, P. Chiacchio, Pnetlab: a tool for the simulation,
analysis and control of discrete event systems based on petri nets, IFAC
Proceedings Volumes 37 (18) (2004) 213–218. \\

[32] E. Kindler, The epnk: an extensible petri net tool for pnml, in: Applications and Theory of Petri Nets: 32nd International Conference, PETRI
NETS 2011, Newcastle, UK, June 20-24, 2011. Proceedings 32, Springer,
2011, pp. 318–327. \\

[33] L. Gomes, F. Moutinho, F. Pereira, Iopt-tools—a web based tool framework for embedded systems controller development using petri nets, in:
2013 23rd International Conference on Field programmable Logic and
Applications, IEEE, 2013, pp. 1–1. \\

[34] T. Freytag, Woped–workflow petri net designer, University of Cooperative Education (2005) 279–282. \\

[35] Y. Thierry-Mieg, Symbolic model-checking using its-tools, in: Tools and
Algorithms for the Construction and Analysis of Systems: 21st International Conference, TACAS 2015, Held as Part of the European Joint
Conferences on Theory and Practice of Software, ETAPS 2015, London,
UK, April 11-18, 2015, Proceedings 21, Springer, 2015, pp. 231–237.\\ 

[36] M. Heiner, M. Herajy, F. Liu, C. Rohr, M. Schwarick, Snoopy–a unifying
petri net tool, in: Application and Theory of Petri Nets: 33rd International Conference, PETRI NETS 2012, Hamburg, Germany, June 25-29,
2012. Proceedings 33, Springer, 2012, pp. 398–407. \\

[37] J. J´ulvez, C. Mahulea, C.-R. V´azquez, Simhpn: A matlab toolbox for
simulation, analysis and design with hybrid petri nets, Nonlinear Analysis: Hybrid Systems 6 (2) (2012) 806–817.\\

[38] F. Pommereau, Snakes: A flexible high-level petri nets library (tool
paper), in: Application and Theory of Petri Nets and Concurrency:
36th International Conference, PETRI NETS 2015, Brussels, Belgium,
June 21-26, 2015, Proceedings 36, Springer, 2015, pp. 254–265. \\

[39] M. Schwarick, C. Rohr, F. Liu, G. Assaf, J. Chodak, M. Heiner, Efficient unfolding of coloured petri nets using interval decision diagrams,
in: Application and Theory of Petri Nets and Concurrency: 41st International Conference, PETRI NETS 2020, Paris, France, June 24–25,
2020, Proceedings 41, Springer, 2020, pp. 324–344. \\

[40] A. Bilgram, P. G. Jensen, T. Pedersen, J. Srba, P. H. Taankvist, Improvements in unfolding of colored petri nets, in: Reachability Problems:
15th International Conference, RP 2021, Liverpool, UK, October 25–27,
2021, Proceedings 15, Springer, 2021, pp. 69–84. \\

[41] E. A. Mahmoud, M. Herajy, I. E. Ziedan, H. I. Shehata, Formal verification confirms the role of p53 protein in cell fate decision mechanism,
Theory in Biosciences 142 (1) (2023) 29–45. \\

[42] A. Bilgram, P. G. Jensen, T. Pedersen, J. Srba, P. H. Taankvist, Methods for efficient unfolding of colored petri nets, Fundamenta Informaticae
189 (2023). \\

[43] S. M. Keating, D. Waltemath, M. K¨onig, F. Zhang, A. Dr¨ager,
C. Chaouiya, F. T. Bergmann, A. Finney, C. S. Gillespie, T. Helikar,
et al., Sbml level 3: an extensible format for the exchange and reuse of
biological models, Molecular systems biology 16 (8) (2020) e9110.\\

[44] J. Pleyer, C. Fleck, Agent-based models in cellular systems, Frontiers in
Physics 10 (2023) 968409. \\

[45]  S. Moretti, V. D. T. Tran, F. Mehl, M. Ibberson, M. Pagni,
Metanetx/mnxref: unified namespace for metabolites and biochemical
reactions in the context of metabolic models, Nucleic acids research
49 (D1) (2021) D570–D574.\\

[46] M. Ostaszewski, A. Niarakis, A. Mazein, I. Kuperstein, R. Phair,
A. Orta-Resendiz, V. Singh, S. S. Aghamiri, M. L. Acencio, E. Glaab,
et al., Covid19 disease map, a computational knowledge repository of
virus–host interaction mechanisms, Molecular systems biology 17 (10)
(2021) e10387.\\

[47] V.-G. Trinh, B. Benhamou, S. Soliman, Trap spaces of boolean networks are conflict-free siphons of their petri net encoding, Theoretical
Computer Science 971 (2023) 114073. \\

[48] M. Weber, E. Kindler, The petri net markup language, in: Petri Net
Technology for Communication-Based Systems: Advances in Petri Nets,
Springer, 2003, pp. 124–144.\\

[49] Petri net markup language (pnml), https://www.pnml.org/, online;
accessed 3rd January 2023 (2015).\\

[50] J. Chodak, M. Heiner, Spike-a command line tool for continuous,
stochastic \& hybrid simulation of (coloured) petri nets, in: Proceedings
of the workshop AWPN, 2018. \\

[51] N. Lanchier, Stochastic modeling, Springer, 2017. \\

[52] S. Dutta, S. Sen, T. Khatun, T. Dutta, S. Tarafdar, Euler number and
percolation threshold on a square lattice with diagonal connection probability and revisiting the island-mainland transition, Frontiers in Physics
7 (2019) 61. \\

[53] X. Feng, Y. Deng, H. W. Bl¨ote, Percolation transitions in two dimensions, Physical Review E 78 (3) (2008) 031136. \\

[54] S. A. Perestrelo, M. C. Gr´acio, N. d. A. Ribeiro, L. M. Lopes, A multiscale network with percolation model to describe the spreading of forest
fires, Mathematics 10 (4) (2022) 588. \\

[55] U.S. Census Bureau, County adjacency file, https://www.census.gov/
geographies/reference-files/2010/geo/county-adjacency.html,
online; accessed May 3, 2023 (August 2023). \\

[56] M. Heiner, M. Schwarick, J.-T. Wegener, Charlie–an extensible petri net
analysis tool, in: Application and Theory of Petri Nets and Concurrency:
36th International Conference, PETRI NETS 2015, Brussels, Belgium,
June 21-26, 2015, Proceedings 36, Springer, 2015, pp. 200–211. \\

[57] K. Schulz, An extension of the snoopy software to process and manage
petri net animations, Bachelor thesis, BTU Cottbus, Dep. of CS (2008).\\

[58] J. Pinto, R. S. Costa, L. Alexandre, J. Ramos, R. Oliveira, Sbml2hyb:
a python interface for sbml compatible hybrid modeling, Bioinformatics
39 (1) (2023) btad044.\\

[59] S. Hwang, M. Chang, Similarity-principle-based machine learning
method for clinical trials and beyond, Statistics in Biopharmaceutical
Research 14 (4) (2022) 511–522.

\bibliographystyle{elsarticle-num} 


\end{document}